\documentclass[preprint,aps,amsmath,showpacs,showkeys,nofootinbib,
superscriptaddress]{revtex4}
\usepackage[dvips]{graphicx}

\begin{document}

\title{\bf Dynamics of $\bar{K}$ and multi-$\bar{K}$ nuclei}

\author{D.~Gazda}
\email{gazda@ujf.cas.cz}
\affiliation{Nuclear Physics Institute, 25068 \v{R}e\v{z}, Czech Republic}

\author{E.~Friedman}
\email{elifried@vms.huji.ac.il}
\affiliation{Racah Institute of Physics, The Hebrew University,
Jerusalem 91904, Israel}

\author{A.~Gal}
\email{avragal@vms.huji.ac.il}
\affiliation{Racah Institute of Physics, The Hebrew University,
Jerusalem 91904, Israel}

\author{J.~Mare\v{s}}
\email{mares@ujf.cas.cz}
\affiliation{Nuclear Physics Institute, 25068 \v{R}e\v{z}, Czech Republic}

\date{\today}

\begin{abstract}
We report on self-consistent calculations of single-$K^-$ nuclear states 
and multi-$\bar{K}$ nuclear states in $^{12}$C, $^{16}$O, $^{40}$Ca and 
$^{208}$Pb within the relativistic mean-field (RMF) approach. 
Gradient terms motivated by the $p$--wave resonance $\Sigma(1385)$ are 
found to play a secondary role for single-$K^-$ nuclear systems where the 
mean-field concept is acceptable. 
Significant contributions from the $\bar{K}N\rightarrow\pi\Lambda$ 
conversion mode, and from the nonmesonic $\bar{K}NN\rightarrow YN$ 
conversion modes which are assumed to follow a $\rho^2$ density dependence, 
are evaluated for the deep binding-energy range of over 100 MeV where the 
decay channel $\bar{K}N\rightarrow\pi\Sigma$ is closed. Altogether we 
obtain $K^-$ total decay widths of 50--100 MeV for binding energies 
exceeding 100 MeV in single-$K^-$ nuclei. 
Multi-$\bar{K}$ nuclear calculations indicate that the binding energy 
per $\bar{K}$ meson saturates upon increasing the number of $\bar{K}$ mesons 
embedded in the nuclear medium. The nuclear and $\bar{K}$ densities increase 
only moderately and are close to saturation, with no indication of any 
kaon-condensation precursor. 

\end{abstract}



\maketitle
\newpage

\section{Introduction}
\label{sec:intro} 
The subject of the present work is the study of $\bar K$ meson interactions with
the nuclear medium. It is closely related to one of the most important, so far 
unresolved problems in hadronic physics, of how hadron masses and interactions
change within the nuclear medium. The in-medium properties of antikaons in 
dense nuclear matter have attracted considerable attention since the 
pioneering work of Kaplan and Nelson on the possibility of kaon condensation 
in dense matter \cite{KNe86,NKa87} and subsequent works offering related 
scenarios in nuclear matter \cite{BLR94,LBM94}. 

The existence of $\Lambda (1405)$, a $\bar{K}N$ quasi-bound state lying about
27~MeV below the $K^- p$ threshold, suggests that the $\bar{K}N$ interaction 
is strongly attractive, as demonstrated first in a vector-meson exchange model 
due to Dalitz {\it et al.}~\cite{DWR67}. This is consistent with low-energy 
$\bar{K}N$ scattering data \cite{Mar81} as well as with the measured level 
shift of the $1s$ state in the kaonic hydrogen atom \cite{Iwa97,Bee05}. 
The $\Lambda(1405)$, as a $K^- p$ quasi-bound state, was also corroborated 
in the J\"{u}lich meson exchange model \cite{GHS90}, where the scalar 
$\sigma$ and vector $\omega$ mesons act jointly to give strong attraction. 
Subsequent chiral SU(3) calculations showed that the $I=0$ coupled-channel 
$\bar{K}N - \pi \Sigma$ interaction is sufficiently attractive to bind the 
$\Lambda(1405)$ \cite{K94,KSW95}. For an update on such calculations see 
Refs.~\cite{BNW05,OPV05,BMN06}. 

The $\bar{K}$--nucleus interaction, too, is strongly attractive, as deduced 
from analyses of strong-interaction level shifts and widths in kaonic atoms 
\cite{FGB93,FGB94,BFG97,FGM99,MFG06,BFr07}. These fits to kaonic-atom data 
are based on phenomenological density dependent optical potentials 
\cite{FGB93,FGB94,BFG97,MFG06,BFr07} or on a relativistic mean-field (RMF) 
approach \cite{FGM99}, yielding strongly attractive $K^-$--nucleus potentials 
of depths 150--200~MeV. For an update see Ref.~\cite{FGa07}. In contrast, 
coupled-channel calculations using $\bar{K}N$ interactions constrained by 
chiral models and fitted to the low-energy $\bar{K}N$ scattering and reaction 
data result in shallower $\bar{K}$--nucleus potentials of depth in the range 
of 100--150~MeV \cite{WKW96}. Imposing a self-consistency requirement on the 
evaluation of the in-medium $\bar{K}$--nucleus potential, yields much 
shallower potentials of depth about 50~MeV \cite{BKE00,ROs00,BGN00,CFG01}. 
Similar results are obtained when the J\"{u}lich meson-exchange model is 
used within a self-consistent coupled-channel calculation \cite{TRP01}. 
A depth of about 80~MeV is indicated by analyzing the enhanced 
near-threshold production of $K^-$ mesons in proton--nucleus collisions, 
in recent experiments by the KaoS Collaboration at GSI~\cite{Sch06} 
(and references cited therein to earlier nucleus-nucleus experiments). 

The $\bar{K}$--nuclear interaction is also strongly absorptive, due 
dominantly to the one-nucleon absorption reactions $\bar{K}N\rightarrow\pi Y$ 
with approximately 100~MeV ($Y=\Sigma$) and 180~MeV ($Y=\Lambda$) energy 
release for the final hyperon $Y$. The strong absorptivity is confirmed by 
fits to kaonic-atom data \cite{BFG97}. 

Considerable interest in this field in recent years has focused on the 
question of possible existence of deeply bound $\bar{K}$--nuclear states, 
and whether such states are sufficiently narrow to allow unique experimental 
identification. Kishimoto~\cite{Kis99}, and Akaishi and Yamazaki 
\cite{AYa99,AYa02}, suggested to look for $\bar{K}$--nuclear states bound by 
over 100~MeV, for which the dominant $\bar{K}N\rightarrow\pi\Sigma$ decay 
channel would become kinematically forbidden. Furthermore, it was suggested 
that multi-$\bar{K}$ high-density nuclear clusters should also exist, 
providing perhaps a precursor stage to kaon condensation \cite{YDA04}. 
Several searches for $\bar{K}$ deeply bound states have been subsequently 
made in KEK \cite{Suz04,Suz05,Iwa07,Kis07}, by the FINUDA collaboration in 
DA$\Phi$NE, Frascati \cite{Agn05} and at the AGS, Brookhaven \cite{Kis05}. 
However, the interpretation of the observed spectra is ambiguous, 
as demonstrated by an alternative explanation of the (allegedly $K^-pp$) 
peak observed in the back-to-back $\Lambda p$ invariant mass distribution 
of $K^-_{\rm stop}$ reactions on $^{6,7}$Li and on $^{12}$C \cite{Agn05} 
in terms of quasi-free $K^-NN$ absorption and final-state 
interaction \cite{MOR06}. 

The theoretical calculations of $\bar{K}$--nuclear bound states may be 
divided into two classes: (i) few-body calculations using a single-channel 
$G$-matrix or AMD methodology~\cite{AYa02,YAk02,DHA04,DWe07}, and 
coupled-channel $\bar{K} NN-\pi \Sigma N$ Faddeev equations~\cite{SGM07,ISa07} 
which agree with Refs.~\cite{AYa02,DWe07} on $K^-pp$ being bound, although 
differing widely on the binding energy and width calculated for this 
lightest possible system; (ii) dynamical RMF calculations \cite{MFG05,ZPL06} 
which take into account the polarization of the nucleus owing to the strongly 
interacting $\bar{K}$ meson, as well as the reduction of phase space 
available for the decay of the deeply bound $\bar{K}$ meson. 
The calculations of Ref.~\cite{MFG06} provide a lower limit of 
$\Gamma_{K^-} \simeq 50$~MeV on the width of nuclear bound states for 
$K^-$ binding energy in the range $B_{K^-}\sim 100-200$~MeV. 

The purpose of the present paper is twofold. In the first part we report on 
dynamical calculations of single-$K^-$ nuclear states within the relativistic 
mean field (RMF) approach. This part of the work deals with three items: 
\begin{itemize} 
\item Effects due to the vector-meson $\rho$ and $\phi$ mean fields which 
were not included in our previous calculations \cite{MFG05} are studied. 
The introduction of the $\rho$-meson mean field allows for a departure from 
$N=Z$ nuclear cores for $\bar K$--nuclear states, and the introduction of the 
$\phi$-meson mean field allows for studying multi-strange $\bar K$--nuclear 
states. 
\item The effect of $p$-wave gradient terms motivated by the $I=1$ 
$\Sigma(1385)$ resonance is studied by extending the RMF coupled equations 
in the simplest form. Although the role of the $\bar{K}N$ $p$-wave interaction 
is marginal near threshold \cite{GNO02}, it might become more important for 
deeply bound antikaons, owing to local variations in the nuclear densities 
induced by the antikaon \cite{WGr05}. 
\item Following our previous work \cite{MFG05}, we explore in more detail and 
rigor the absorptive part of the optical potential in the energy region where 
the dominant decay channel $\bar{K}N\rightarrow\pi\Sigma$ is closed. This is 
done by incorporating for the first time the $\bar{K}N\rightarrow\pi\Lambda$ 
channel, with threshold some 80~MeV below the $\pi\Sigma$ threshold, and by 
considering a $\rho^2$ density dependence for the two-nucleon absorption 
modes $\bar{K}NN\rightarrow YN$. The $\rho^2$ dependence is more appropriate 
for the description of the two-nucleon nature of these absorption modes. 
\end{itemize} 

In the second part of this work we explore within the RMF methodology deeply
bound multi-$\bar{K}$ nuclear states, in order to study the behavior of the
nuclear medium under the influence of increasing strangeness.  The issue here is
whether or not the binding energy per $\bar{K}$ meson of multi-$\bar{K}$ nuclear
states increases significantly upon adding a large number of $\bar{K}$ mesons,
so that $\bar{K}$ mesons provide the physical degrees of freedom for self-bound
strange hadronic systems. Kaon condensation in nuclear matter would occur beyond
some threshold value of strangeness, if the binding energy per $\bar{K}$ meson
exceeds the combination $m_K + \mu_N - m_{\Lambda} \gtrsim 320$~MeV (in this
paper we use $\hbar=c=1$), where $\mu_N$ is the nucleon chemical potential. In
such a case, $\Lambda$, $\Sigma$ and $\Xi$ hyperons would no longer combine
macroscopically with nucleons to compose the more conventional kaon-free form of
strange hadronic matter \cite{BGa00}. In neutron stars, the binding energy per
$\bar{K}$ meson necessary for the onset of kaon condensation is given by $m_K -
\mu_{e^-}$, where $\mu_{e^-}$ is the electron chemical potential (generally
accepted to assume values of $\mu_{e^-} \lesssim 200$~MeV). The RMF approach was
first applied to the study of kaon condensation in the mid 1990s, originally
without considering a possible interplay with hyperons \cite{SGM94,BRh96} and
then with hyperons included \cite{KPE95,SMi96}.  This approach was also used in
Ref.~\cite{PBG00} to consider $\bar{K}^0$ condensation in addition to $K^-$
condensation in neutron stars.  Recent calculations offer a wide range of
interesting precursor phenomena to kaon condensation in both hadronic and quark
sectors (see Refs.~\cite{MMT05,BLP06,RHH07,KKM07,Mut07} and previous work 
cited therein). The present calculations may shed some light on the likelihood 
of a kaon-condensation scenario in nuclear matter. 

The RMF methodology, including the extension to absorptive processes and 
$p$-wave gradient interaction terms, and to multi-$\bar{K}$ nuclear states 
is discussed in Sec.~\ref{sec:model}. Results of calculations for 
a representative set of nuclear cores across the periodic table are 
presented and discussed in Sec.~\ref{sec:results}. Section~\ref{sec:concl} 
summarizes the new results of the present work, along with conclusions and 
outlook.
\section{Model} 
\label{sec:model}
In the present work, $\bar{K}$-nuclear states are studied within the 
theoretical framework of the relativistic mean field (RMF) approach applied 
to a system of nucleons and one or more $\bar{K}$ mesons. The interaction 
among hadrons is mediated by the exchange of scalar and vector meson fields. 
The standard RMF Lagrangian density ${\cal L}_N$ describing the nucleonic 
sector is specified in Sec.~\ref{sec:results}. The (anti)kaonic sector is 
incorporated by adding to ${\cal L}_N$ the Lagrangian density ${\cal L}_K$:
\begin{equation}
{\mathcal L}_{K} =
\left( {\mathcal D}_\mu K \right)^\dagger
\left( {\mathcal D}^{\,\mu} K \right)
-m_K^2 K^\dagger K
+g_{\sigma K} m_K K^\dagger K \, \sigma
\:,
\end{equation}
with
\begin{equation}
K=\left(
\begin{array}{c}
K^+ \\ K^0
\end{array}
\right)
\qquad
K^\dagger=\left(
\begin{array}{cc}
K^-, & \bar{K}^0
\end{array}
\right)
\end{equation}
and the covariant derivative ${\cal D}_\mu$ given by:
\begin{equation}
{\mathcal D}_\mu \equiv
\partial_\mu
+ {\rm i}\, g_{\omega K}\, \omega_\mu
+ {\rm i}\, g_{\rho K}\, \vec{\tau} \cdot \vec{\rho}_\mu
+ {\rm i}\, g_{\phi K}\, \phi_\mu
+ {\rm i}\, e\, \textstyle\frac{1}{2}\displaystyle(1+\tau_3) A_\mu
\:.
\end{equation}
This particular choice of interaction scheme leads to the coupling of the 
vector meson fields to conserved currents. The conserved (Noether) current 
associated with the kaonic field is obtained from the invariance of 
${\mathcal L}_K$ under global phase transformation. Using 
\begin{equation}
K \rightarrow e^{{\rm i} \lambda}K\quad\Rightarrow\quad
\delta {\mathcal L}_K = 0 \quad\Rightarrow\quad
\partial_\mu\, j^{\mu}_K=
\partial_\mu \left[ \frac{\delta {\mathcal L}_K}{\delta \left( \partial_\mu K 
\right)}\,\delta K +
\delta K^\dagger\frac{\delta {\mathcal L}_K}{\delta \left( \partial_\mu
K^\dagger \right)} \right]=0 \:,
\end{equation}
one obtains a conserved vector current whose vacuum (represented by 
filled shells of nucleons and $\kappa$ $\bar{K}$ mesons) expectation value 
transforms to the expression for $\rho_{K^-}$ given in Eq.~(\ref{eq:kdens}) 
below. The standard variational principle yields equations of motion for all 
field operators. The meson field operators and source terms are then replaced 
by their expectation values, according to the mean-field approximation. 
For simplicity, we limit discussion in this section to nuclear systems with 
$K^-$ mesons. The generalization to nuclear systems with ${\bar K}^0$ mesons 
is straightforward. 

Whereas the Dirac equation for nucleons is not explicitly affected by 
the addition of ${\cal L}_K$, the presence of $K^-$ mesons induces additional 
source terms in the equations of motion for the meson (mean) fields:
\begin{eqnarray}
\nonumber
(-\nabla^2 + m_\sigma^2)\sigma_0 &=&
+ g_{\sigma N} \rho_s +g_2 \sigma_0^2 -g_3 \sigma_0^3
+ g_{\sigma K} m_K K^- K^+
\\ \nonumber
(-\nabla^2 + m_\omega^2) \omega_0 &=&
+ g_{\omega N} \rho_v
- g_{\omega K} \rho_{K^-}
\\ \label{eq:mkg}
(\,-\nabla^2 + m_\rho^2) \rho_0 &=&
+ g_{\rho N} \rho_3 - g_{\rho K} \rho_{K^-}
\\ \nonumber
(-\nabla^2 + m_\phi^2)\phi_0 &=&
- g_{\phi K} \rho_{K^-}
\\ \nonumber
-\nabla^2 A_0 &=& + e\, \rho_p - e\, \rho_{K^-}
\, ,
\end{eqnarray}
with $\rho_s$, $\rho_v$ and $\rho_3$ denoting the scalar, vector and 
isovector nuclear densities, respectively, and with $\rho_p$ denoting 
the proton density. The $K^-$ density $\rho_{K^-}$ is 
given by:
%
\begin{equation} 
\label{eq:kdens} 
\rho_{K^-}=2
(E_{K^-}+g_{\omega K}\,\omega_0+g_{\rho K}\,\rho_0+g_{\phi K}\,\phi_0+e\, A_0)
K^-K^+ \, ,
\qquad \int {\rm d}^3 x\, \rho_{K^-} = \kappa
\:,
\end{equation}
where $E_{K^-}={\rm i}\,\partial_t K^-$. The density $\rho_{K^-}$ is 
normalized to the number of antikaons $\kappa$ in a multi-$K^-$ system.

The Klein--Gordon (KG) equation of motion for the $K^-$ meson obtained from 
the RMF model acquires the form: 
\begin{equation}
\label{eq:Kkg}
[-\nabla^2-E_{K^-}^2 +m_K^2 + {\rm Re}\,\Pi_{K^-} \,]K^-=0
\:, 
\end{equation}
where the $K^-$ self-energy term is given by:
\begin{eqnarray}
{\rm Re}\,\Pi_{K^-}=
&-&g_{\sigma K}m_K\sigma_0
-2E_{K^-}(g_{\omega K}\omega_0+g_{\rho K}\rho_0+g_{\phi K}\phi_0+eA_0)
\\ \nonumber
&-&(g_{\omega K}\omega_0+g_{\rho K}\rho_0+g_{\phi K}\phi_0+eA_0)^2
\:.
\end{eqnarray}
This implies a $K^-$ effective mass $m_K^*$ of the form 
$m_K^{*2}=m_K^2-g_{\sigma K}m_K \sigma_0$, in contrast to another possible 
choice \cite{PBG00} $m_K^*=m_K-g_{\sigma K}\sigma_0$. Qualitatively, our 
results are insensitive to this difference and the conclusions of the 
present study hold also for the other choice.

Assuming that all the $K^-$ mesons occupy the same energy level, 
the total binding energy of the combined $\kappa K^-$--nuclear system 
$B(A,Z,\kappa K^-)$ 
has the form:
\begin{eqnarray}
B(A,Z,\kappa K^-)&=&\textstyle\sum_{i=1}^A\displaystyle B_i^{\rm sp} + 
\kappa\,B_{K^-}^{\rm sp} \\ \nonumber
&-&\frac{1}{2}\int {\rm d}^3\,x(-g_{\sigma N}\,\sigma_0\rho_s+g_{\omega
N}\,\omega_0 \rho_v+g_{\rho N}\,\rho_0\rho_3+e\,A_0\rho_p)
\\ \nonumber
&-&\frac{1}{2}\int {\rm d}^3\,x
(-\textstyle\frac{1}{3}\displaystyle
g_2\,\sigma_0^3-\textstyle\frac{1}{2}\displaystyle
g_3\,\sigma_0^4)
\\ \nonumber
&+&\frac{1}{2}\int {\rm d}^3\,x [(g_{\omega K}\,\omega_0+g_{\rho
K}\,\rho_0+g_{\phi K}\,\phi_0+e\,A_0)\rho_{K^-}+g_{\sigma K}m_K \sigma_0 
K^- K^+] \:,
\end{eqnarray}
with $B_i^{\rm sp}=m_N -E_i$ and $B_{K^-}^{\rm sp}=m_K -E_{K^-}$, where 
$E_i$ and $E_{K^-}$ are the nucleon and $K^-$ single particle energies, 
respectively. From this expression, it is evident that the $K^-$ binding 
energy $B_{K^-}=B[A,Z,\kappa {K^-}]-B[A,Z,(\kappa-1){K^-}]$ contains, 
in addition to $B_{K^-}^{\rm sp}$, mean field contributions representing 
part of the rearrangement energy of the nuclear core.

\subsection{P-wave contributions} 
\label{sec:pwave} 
To study the role of $p$ waves in the $\bar{K}N$ interaction, we have 
extended the $K^-$ self energy $\Pi_{K^-}$ by adding a phenomenological 
isoscalar $p$--wave potential: 
\begin{equation}
\Pi^{(P)}_{K^-}=4\pi\left(1+\frac{E_{K^-}}{m_N}\right)^{-1} c_0 
(\nabla \rho_v ) \cdot \nabla \:, 
\label{eq:pwave} 
\end{equation}
where $c_0 = (c_p + c_n)/2 \approx 3c_p/2$ is an energy-dependent strength 
parameter dominated by the contribution of the $\Sigma(1385)$ $p$-wave 
resonance \cite{Wei07}. 
Calculations were done for $K^-$--nuclear states bound by about 
100 MeV, corresponding roughly to $\sqrt{s_{\bar K N}} = 1330$~MeV, 
namely about 55 MeV below the $\Sigma(1385)$ resonance, where $c_0$ 
is positive (attractive) and nearly real: 
\begin{equation} 
c_0(1330~{\rm MeV}) = {\frac{3}{2}}~c_p(1330~{\rm MeV})=0.186~{\rm fm}^3 \:, 
\label{eq:c0} 
\end{equation} 
in agreement with the plot of $c_p$ in Fig.~2 of Ref.~\cite{DWe07}. 

\subsection{Absorptive contributions}
\label{sec:abs} 
Having dealt with the nuclear binding energy of $K^-$ mesons, in the next 
step we consider $\bar{K}$ absorption in the nuclear medium, in order to 
evaluate the $K^-$ decay width $\Gamma_{K^-}$. In our model, 
this is done by allowing the self energy $\Pi_{K^-}$ to become
complex and replacing $E_{K^-}\rightarrow~E_{K^-}-{\rm i}\Gamma_{K^-}/2$. 
Since the imaginary part of the self energy is not addressed by the 
traditional RMF approach, ${\rm Im}\,\Pi_{K^-}$ was taken from optical 
model phenomenology. We follow Ref.~\cite{MFG05} taking the optical 
potential imaginary-part depth from fits to $K^-$ atomic data, while the 
nuclear density is treated as a dynamical quantity in these self-consistent 
calculations. Once the antikaon is embedded in the nuclear medium, the 
attractive $\bar{K}N$ interaction compresses the nuclear core, 
thus increasing the nuclear density which leads to an increased 
$\bar{K}$ width. On the other hand, the phase space available for decay 
products is reduced due to the binding energy of the $\bar{K}$ meson, 
particularly in the case of $\bar{K}$ deeply bound states. 
To accomplish this reduction, suppression factors multiplying 
${\rm Im}\,\Pi_{K^-}$ were introduced, explicitly considering $\bar{K}$ 
binding energy for the initial decaying state and assuming two-body 
final state kinematics.

We first consider absorption on a single nucleon, leading to the following 
pionic decay modes: 
\begin{equation}
\bar{K}N \rightarrow \pi \Sigma,\,\pi \Lambda \quad (70\%,\,10\%) \:,
\end{equation}
with thresholds about 100 MeV and 180 MeV, respectively, below the $\bar{K}N$ 
total mass. The numbers in parentheses give approximately the branching ratios 
known from bubble chamber and emulsion experiments \cite{WSW77}. 
The corresponding single-nucleon absorptive contribution to the optical 
potential is given in leading approximation by: 
\begin{equation}
{\rm Im}\,\Pi_{K^-}^{(1)}=(0.7f_{1\Sigma}+0.1f_{1\Lambda})\,W_0\,\rho_v(r) \:,
\label{eq:1Nabs} 
\end{equation}
where 
$W_0$ is taken from kaonic atom fits and the phase-space suppression factors 
$f_{1Y}$ ($Y=\Sigma,\Lambda$) are given by:
\begin{equation}
f_{1Y}=
\frac{M_{01}^3}{M_1^3}
\sqrt{
\frac
{[M_1^2-(m_\pi+m_Y)^2][M_1^2-(m_Y-m_\pi)^2]}
{[M_{01}^2-(m_\pi+m_Y)^2][M_{01}^2-(m_Y-m_\pi)^2]}
}
\,\Theta(M_1-m_\pi-m_Y)
\:,
\end{equation}
with $M_{01}=m_K+m_N$ and $M_1=M_{01}-B_{K^-}$. 

Absorption on two nucleons leads to non-pionic decay modes 
\begin{equation}
\bar{K}NN \rightarrow YN \quad (20\%)
\:,
\end{equation}
with thresholds about $m_\pi\simeq 140$ MeV lower than the corresponding 
pionic decay mode thresholds. 
Since the non-pionic modes are heavily dominated by the $\Sigma N$ final 
state, the $\Lambda N$ channel was not considered in the present work 
and we focused attention primarily on a quadratic density dependence of the 
$\Sigma N$ final-state contribution to the absorptive part of the optical 
potential. A quadratic density dependence for two-nucleon absorption processes 
has been successfully used in studies of pionic atoms~\cite{BFG97,FGa07}. 
The $\bar{K}$ two-nucleon absorptive part of the optical potential is 
given by:
\begin{equation}
{\rm Im}\,\Pi_{K^-}^{(2)}=0.2\,f_{2\Sigma}\,W_0\,\rho_v^2 (r) / \rho_0 \:,
\label{eq:2Nabs} 
\end{equation}
where the factor 0.2 represents the approximately $20\%$ branching ratio for 
two-nucleon absorption from rest \cite{WSW77} and $\rho_0\sim 0.16$~fm$^{-3}$ 
is an $A$-dependent central nuclear density used for properly normalizing the 
two-nucleon absorption strength with respect to the one-nucleon absorption 
strength. The phase-space suppression factor $f_{2\Sigma}$ has the form:
\begin{equation}
f_{2\Sigma}=
\frac{M_{02}^3}{M_2^3}
\sqrt{
\frac
{[M_2^2-(m_N+m_\Sigma)^2][M_2^2-(m_\Sigma-m_N)^2]}
{[M_{02}^2-(m_N+m_\Sigma)^2][M_{02}^2-(m_\Sigma-m_N)^2]}
}
\,\Theta(M_2-m_\Sigma-m_N)
\:,
\end{equation}
with $M_{02}=m_K+2 m_N$ and $M_2=M_{02}-B_{K^-}$.

The set of coupled equations containing the Dirac equation for nucleons, 
the KG equations (\ref{eq:mkg}) and (\ref{eq:Kkg}) for the meson mean fields 
and for antikaons, respectively, was solved self-consistently using 
an iterative procedure.

\section{Results and discussion}
\label{sec:results} 
Calculations of $\bar K$-nuclear states in $^{12}$C, $^{16}$O, $^{40}$Ca, and 
$^{208}$Pb were performed, using both the linear (HS) \cite{HSe81} and 
non-linear (NL-SH) \cite{SNR93} parameterizations of the nucleonic Lagrangian 
density ${\mathcal L}_N$. These RMF parameterizations give quite different 
estimates of nuclear properties. In particular, the non-linear models yield 
generally lower values of the nuclear incompressibility.
Therefore, stronger polarization effects in these models owing to the presence 
of $\bar K$ meson(s) are anticipated, in comparison with the linear models.

The (anti)kaon coupling constants to the meson fields were chosen as follows: 
The coupling constant $g_{\omega K}$ was given a reference value 
$g_{\omega K}^0=\,(1/3) g_{\omega N}$ following the simple quark model.
The reference value for $g_{\sigma K}$ was taken then from fits to kaonic 
atom data, which yielded $g_{\sigma K}^0=0.2\,g_{\sigma N}$ for the linear 
and $g_{\sigma K}^0=0.233\,g_{\sigma N}$ for the non-linear parameterizations 
of ${\mathcal L}_N$ \cite{FGM99}. Finally, the coupling constants 
$g_{\rho K}$ and $g_{\phi K}$ were adopted from the $SU(3)$ relations: 
$\sqrt{2}\,g_{\phi K}=2\,g_{\rho K}=\,g_{\rho \pi}=6.04$ \cite{SMi96}. 

The $SU(3)$ relation $2g_{\omega K} = g_{\rho \pi}$ was not imposed on
$g_{\omega K}$ since its value was varied in the calculations, along with
varying $g_{\sigma K}$, in order to scan over different values of $\bar{K}$
binding energies. Thus a particular way of varying these coupling constants away
from their `reference' values was used.  Starting from $g_{i
K}\equiv\alpha_i\,g_{i K}^0=0\quad(i=\sigma ,\omega)$, we first scaled up
$\alpha_\omega$ from the value required for the onset of binding all the way to
$\alpha_\omega = 1$, corresponding to $g_{\omega K}=g_{\omega K}^0$.  Then, for
$\alpha_\omega=1$, we scaled up $\alpha_\sigma$ from 0 to 1 corresponding to
$g_{\sigma K}=g_{\sigma K}^0$, and finally we again scaled $\alpha_\omega$ from
1 upwards until the binding energy value of $B_{\bar K}\simeq200$~MeV was
reached. Generally, similar results and conclusions are reached if different
scanning procedures are applied. We comment below, for multi-$\bar{K}$ nuclei,
when this is no longer the case.

\subsection{Single-$K^-$ nuclei} 
\label{sec:single} 
In the first part of this work, single-$K^-$ nuclear states were studied. 
We verified that the interaction generated by the $\rho$-meson mean field 
has a small effect on the $K^-$ binding energy $B_{K^-}$ and on the width 
$\Gamma_{K^-}$, for $B_{K^-}\lesssim 200$~MeV and for all the RMF 
parameterizations considered in the present work. This interaction acts 
repulsively on a $K^-$ meson, producing a small decrease of $B_{K^-}$, 
less than 5~MeV in the case of $^{208}$Pb where the most significant effect 
is anticipated due to the large excess of neutrons. The effect of the 
$\rho$ meson field on the $K^-$ decay width is even smaller, except in the 
region 60~MeV~$ \lesssim B_{K^-} \lesssim $~100~MeV where the phase-space 
suppression factor $f_{1 \Sigma}$ varies rapidly and, hence, $\Gamma_{K^-}$ 
increases by approximately 10~MeV. 

\begin{figure} 
\includegraphics[width=12cm]{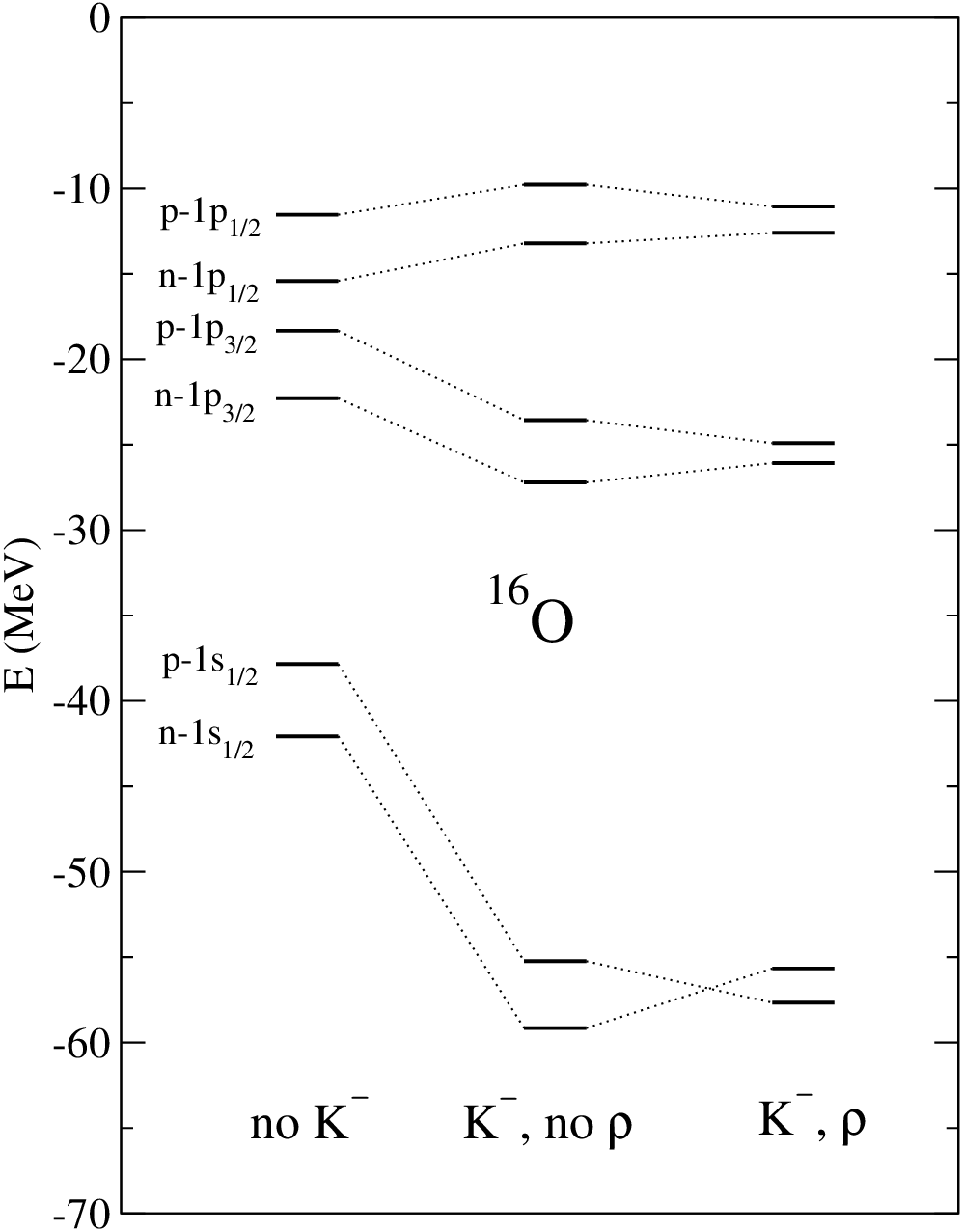}
\caption{Nucleon single-particle energies with respect to the nucleon mass in $^{16}$O (left spectrum) and in 
$_{K^-}^{16}$O (middle and right spectra) for $B_{K^-}=100$ MeV, 
with and without coupling the $K^-$ meson to the $\rho$-meson mean field, 
using the NL-SH RMF model.}
\label{fig:rhomeson}
\end{figure}

Figure \ref{fig:rhomeson} shows the effect of the $\rho K^-$ coupling 
on the nucleon single-particle energies in $^{16}$O. The left-hand spectrum 
shows the nucleon single-particle energies in the absence of $K^-$ mesons, 
using the NL-SH model. The middle spectrum displays the rearrangement of these 
single-particle energies caused by a $K^-$ meson bound by 100 MeV, with no 
$\rho K^-$ coupling. The most pronounced effect is observed for the 
$1s_{1/2}$ nucleon states, which become significantly more bound in the 
presence of a $1s$ $K^-$ meson. The right-hand spectrum displays further 
modification of the nucleon single-particle energies due to the $\rho K^-$ 
coupling. It is seen that the isovector $\rho K^-$ interaction reverses 
the order of the $1s_{1/2}$ proton and neutron energy levels, determined in 
the absence of $\rho K^-$ coupling by the Coulomb interaction. 
This reversal is due to the $\rho K^-$ coupling acting against the 
Coulomb interaction. 

The interaction generated by the $\phi$-meson mean field reduces the $K^-$ 
binding energy in systems with more than one $K^-$ meson, as it mediates 
repulsive interaction exclusively among strange particles. Generally, for the 
parameterizations and nuclei studied, the effect of the $\phi$-meson repulsion 
increases with $B_{K^-}$, owing to the increased central $K^-$ density. 
It amounts to several MeV for binding energies $B_{K^-}\lesssim 200$ MeV. 

In the next step, considering the $K^-$ decay modes discussed in the 
previous section, we calculated the corresponding widths of $K^-$-nuclear 
bound states. In particular, we considered the one-nucleon absorption mode 
$\bar K N \to \pi\Lambda$, in addition to the dominant 
$\bar K N \to \pi\Sigma$ mode studied in our recent work \cite{MFG05}, 
and also both $\rho$ and $\rho^2$ density dependencies of the two-nucleon 
absorption mode $\bar K NN \to \Sigma N$. 

\begin{figure}
\includegraphics[width=12cm]{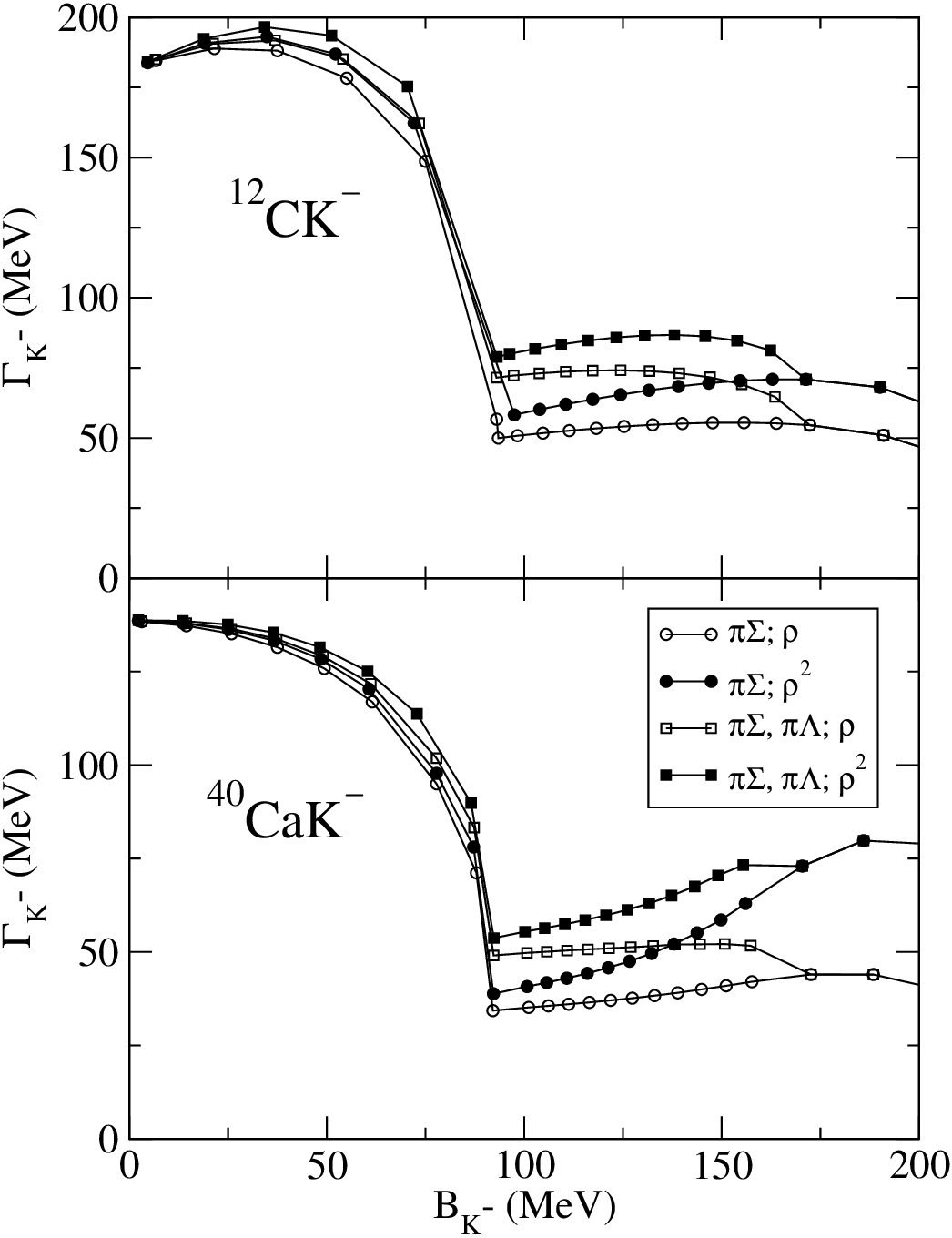}
\caption{Widths of the $1s$ $K^-$-nuclear state in $^{12}_{K^-}$C (top panel) 
and in $^{40}_{K^-}$Ca (bottom panel) as a function of the $K^-$ binding energy, 
for absorption through ${\bar K}N \rightarrow \pi\Sigma$, with and without 
${\bar K}N \rightarrow \pi\Lambda$, and assuming $\rho$ or $\rho^2$ 
dependence for ${\bar K}NN \rightarrow {\Sigma}N$ 
(using the NL-SH RMF model).} 
\label{fig:GC}
\end{figure}

\begin{figure}
\includegraphics[width=12.cm]{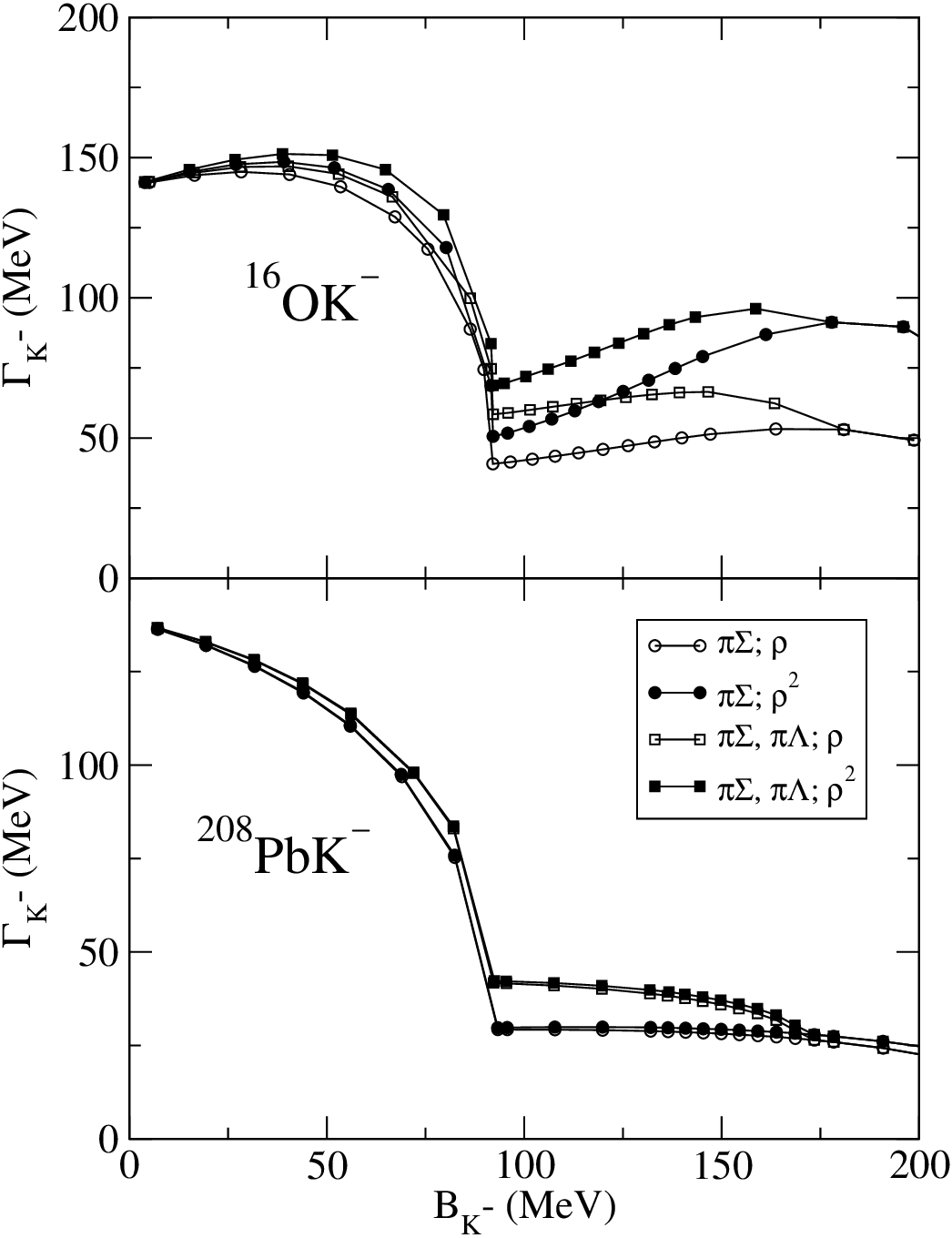}
\caption{Widths of the $1s$ $K^-$-nuclear state in $^{16}_{K^-}$O using the 
NL-SH RMF model (top panel), and in $^{208}_{K^-}$Pb using the HS RMF model 
(bottom panel), as a function of the $K^-$ binding energy for various 
combinations of and assumptions on the $K^-$ absorption modes as in 
Fig.\ \protect{\ref{fig:GC}}.} 
\label{fig:GOPb}
\end{figure}

Figure \ref{fig:GC} shows the calculated width $\Gamma_{K^-}$ as a function of 
the binding energy $B_{K^-}$ for the $K^-$ $1s$ state in $^{12}_{K^-}$C (top) 
and in $^{40}_{K^-}$Ca (bottom), using the nonlinear model NL-SH. 
The effect of allowing the $\pi\Lambda$ decay mode ($10\%$) to share 
alongside with $\pi\Sigma$ ($70\%$) the one-nucleon absorption strength 
is shown by squares, compared to circles for the $\pi\Sigma$ mode alone 
($80\%$). In each of the two groups of curves shown (one for $\rho$ and 
the other one for $\rho^2$ density dependence of the two-nucleon mode) the 
inclusion of the secondary $\pi\Lambda$ decay mode contributes approximately 
20 MeV in $^{12}_{K^-}$C and 15 MeV in $^{40}_{K^-}$Ca to the width in the 
region of binding energies $B_{K^-}$ between 100 and 160 MeV. 
As for the density dependence of the two-nucleon absorption mode, 
the widths calculated assuming $\rho$ and $\rho^2$ density dependence 
are denoted by open and solid symbols, respectively. 
Assuming $\rho^2$ instead of $\rho$ density dependence, it leads to increased 
widths of the $1s$ $K^-$-nuclear states, as demonstrated for $^{12}_{K^-}$C 
and $^{40}_{K^-}$Ca in Fig.~\ref{fig:GC}, and for $^{16}_{K^-}$O in 
Fig.~\ref{fig:GOPb}. The effect of the $\rho^2$ dependence of the 
$2N$-absorption mode clearly grows with $B_{K^-}$ as a consequence of the 
enhanced central nuclear density $\rho_N$. While for $B_{K^-}\lesssim 100$~MeV 
it is less than 10~MeV, for $B_{K^-}\gtrsim 150$~MeV it amounts to 
about 20~MeV in C and as much as about 30~MeV in Ca. 

\begin{figure}
\includegraphics[width=14cm]{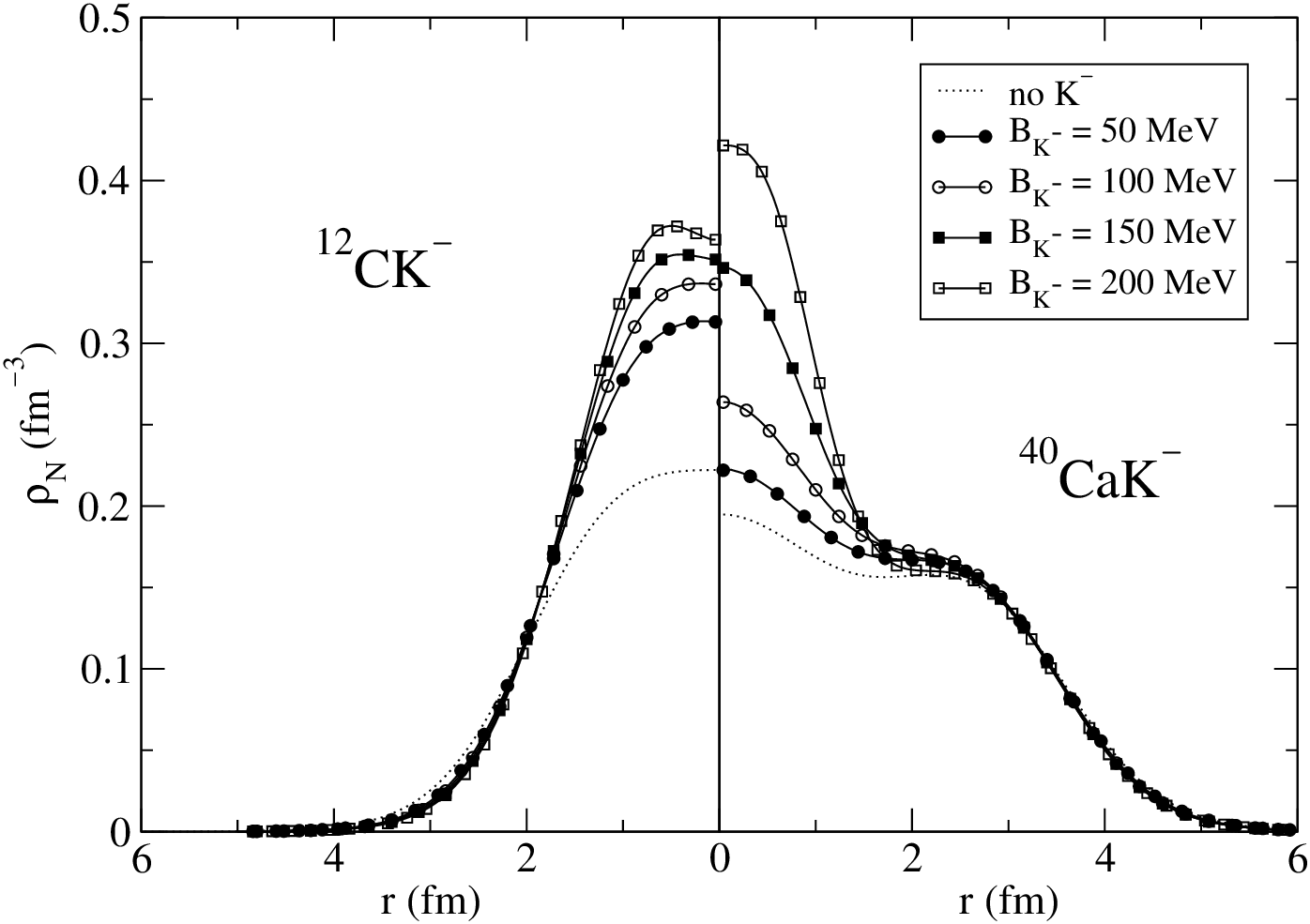}
\caption{Nuclear density $\rho_N$ of $^{12}_{K^-}$C (left panel) and 
$^{40}_{K^-}$Ca (right panel) for several $1s$ $K^-$-nuclear states with 
specified binding energy, using the NL-SH RMF model. The dotted curves denote 
the corresponding nuclear density in the absence of the $K^-$ meson.} 
\label{fig:CaC}
\end{figure}

Figure~\ref{fig:GOPb} shows the widths $\Gamma_{K^-}$ in $^{16}_{K^-}$O 
for the nonlinear model NL-SH (top) and in $^{208}_{K^-}$Pb for the linear 
model HS (bottom). As in the previous figure, switching on the $\pi\Lambda$ 
decay channel adds further conversion width to $K^-$-nuclear states. In the 
range $B_{K^-} \simeq 100 - 160$~MeV the width $\Gamma_{K^-}$ increases by 
about 20~MeV. The $\pi\Lambda$ conversion mode disappears at 
$B_{K^-} \simeq 175$~MeV. The effect of the $\pi\Lambda$ absorption channel 
is almost uniform for both nonlinear (NL-SH in Fig.\ \ref{fig:GC} and 
Fig.\ \ref{fig:GOPb}, top) and linear (HS in Fig.\ \ref{fig:GOPb}, bottom) 
parameterizations in all nuclei under consideration. On the other hand, the 
widths calculated assuming $\rho^2$ dependence for the two-nucleon 
absorption mode exhibit strong sensitivity to the type of RMF model applied 
and to the nucleus considered (via the nuclear density  $\rho_N$). 
In $^{208}_{K^-}$Pb, there is almost no difference between the widths 
$\Gamma_{K^-}$ calculated using $\rho$ or $\rho^2$ dependence. It was found 
that nonlinear parameterizations, represented here by NL-SH, produce larger 
increase of the width $\Gamma_{K^-}$ owing to a $\rho^2$ density dependence 
of the $2N$ $K^-$ absorption than linear models do, as could be anticipated 
from the considerably lower incompressibilities predicted by nonlinear models. 
It is to be noted that the particularly large widths $\Gamma_{K^-}$ in 
$^{40}_{K^-}$Ca for $B_{K^-}\gtrsim 150$~MeV are due to a more significant 
increase of the central nuclear density in $^{40}_{K^-}$Ca than in 
$^{12}_{K^-}$C within the NL-SH model, see Fig.\ \ref{fig:CaC}. 

Figure \ref{fig:CaC} demonstrates that the effect of nuclear compression, 
as evidenced by the increase of the nuclear density $\rho_N$ upon increasing 
the binding energy $B_{K^-}$ of the $1s$ state, is limited to relatively 
small radii, $r \lesssim 1.5$~fm. Whereas in as light a nucleus as C this 
region constitutes most of the nucleus, it is only a fraction of the nuclear 
volume in medium-weight nuclei such as Ca and in heavier nuclei (not 
shown in the figure).  

\begin{figure} 
\includegraphics[width=12cm]{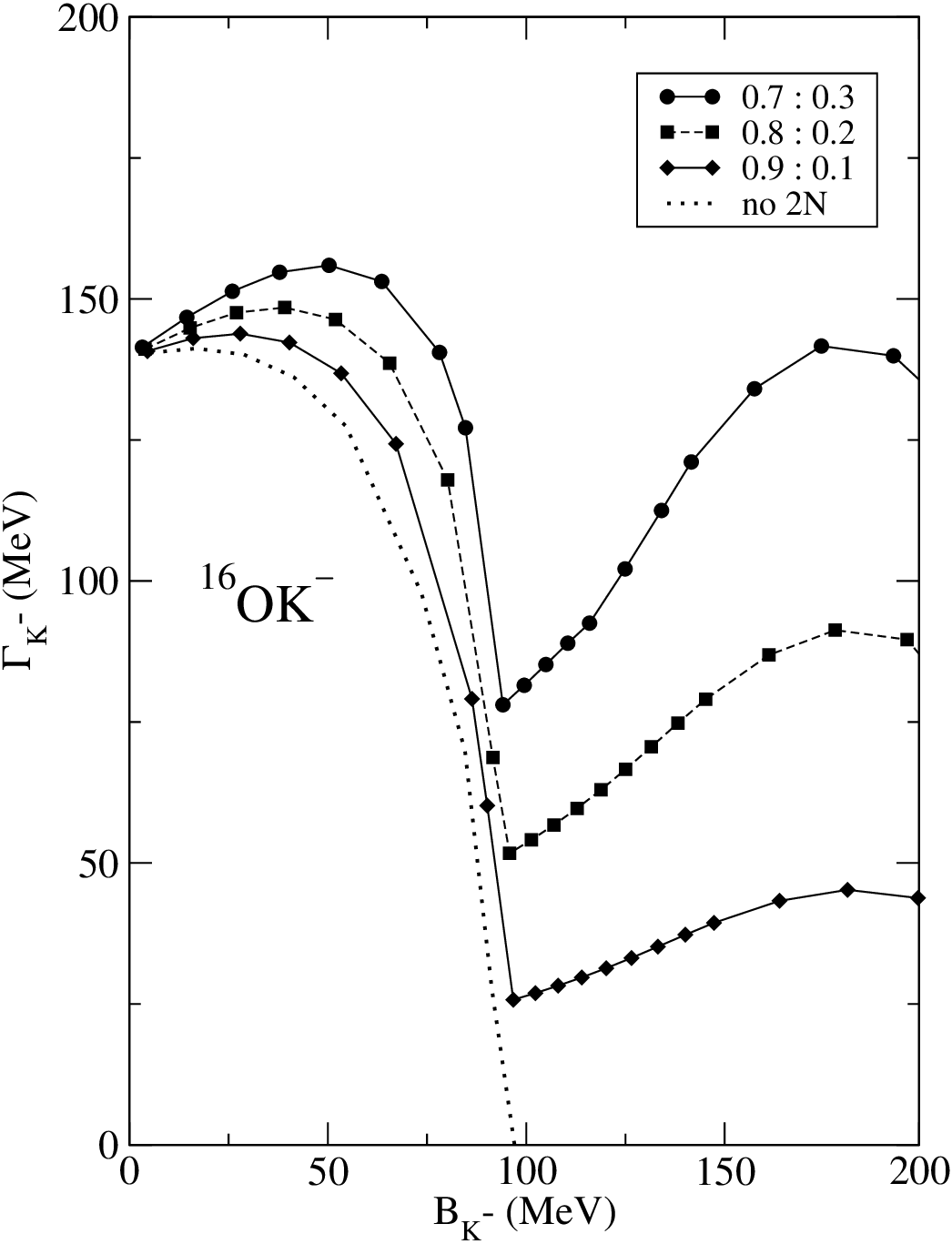} 
\caption{Widths of the $1s$ $K^-$-nuclear state in $^{16}_{K^-}$O for 
various absorption branching ratios 
$\bar{K}N\rightarrow\pi\Sigma:\bar{K}NN\rightarrow\Sigma N$, using the 
NL-SH RMF model and $\rho^2$ dependence for the $2N$-absorption channel. 
The dotted curve stands for the decay widths in the absence of  
$2N$-absorption.} 
\label{fig:bratios}
\end{figure}

We also studied another possible source of uncertainty for the calculated 
$K^-$ decay width, namely the dependence on the branching ratios assumed for 
the various conversion modes. Figure~\ref{fig:bratios} shows the $K^-$ 
decay width $\Gamma_{K^-}$ as a function of the $K^-$ binding energy $B_{K^-}$ 
in $^{16}_{K^-}$O for the NL-SH model, assuming a $\rho^2$ dependence of 
the $2N$-conversion mode. The branching ratios of the decay modes, 
$\bar{K}N\rightarrow\pi\Sigma : \bar{K}NN\rightarrow \Sigma N$, are varied 
from $0.7:0.3$ (circles) to $0.8:0.2$ (squares) and to $0.9:0.1$ (diamonds). 
The dotted curve represents the decay widths calculated when the 
$2N$-absorption modes are neglected altogether. It is shown that varying  
the $K^-$ absorption branching ratios by $\pm 0.1$ away from the commonly 
used value $0.8:0.2$ alters the $K^-$ decay width $\Gamma_{K^-}$ by less than 
10~MeV for binding energies $B_{K^-}\lesssim 90$~MeV. More remarkable is the 
effect in the region of $B_{K^-}\gtrsim 90$~MeV, where the dispersion reaches 
values of approximately 50~MeV. These results further point out to the 
delicacy of the estimates for the $K^-$ decay widths in that region of 
binding energies. It is worth noting that the $0.8:0.2$ `canonical' ratio 
is used here in a rather conservative way, implicitly assuming that it is 
effective for capture in the nuclear central-density region 
[see Eq.~(\ref{eq:2Nabs}) for Im~$\Pi_{K^-}^{(2)}$], whereas capture from 
rest in bubble-chamber and emulsion experiments \cite{WSW77} is likely 
to occur in lower-density regions. Therefore, the contribution to 
the $K^-$ decay widths due to Im~$\Pi_{K^-}^{(2)}$ could be larger than 
estimated by adopting the $0.8:0.2$ `canonical' ratio. The ambiguities 
involved in evaluating the contribution of Im~$\Pi_{K^-}^{(2)}$ have been 
recently discussed by Yamagata and Hirenzaki~\cite{YHi07}. 

The last item studied for single-$K^-$ nuclear states was the effect of 
a $p$-wave $K^-$-nucleus interaction $\Pi_{K^-}^{(P)}$ [Eq.~(\ref{eq:pwave})]. 
Table~\ref{tab:1s} demonstrates the effects of this interaction, 
with a strength parameter $c_0$ given by Eq.~(\ref{eq:c0}) for 
a nominal value of $B_{K^-}=100$~MeV. Shown are the 
$K^-$ binding energy $B_{K^-}$, the single-particle binding energy 
$B_{K^-}^{\rm sp}$ and the decay width $\Gamma_{K^-}$, calculated for $1s$ 
$K^-$--nuclear states using the NL-SH parameterization. 
The results using the linear HS model are almost the same. 
The calculations excluding the $p$-wave $K^-$ interaction are denoted by S, 
while those including the $p$-wave interaction are denoted by S+P. 
It is seen that the introduction of the $p$-wave interaction leads to an 
increase of the binding energy by approximately 13~MeV in $^{12}_{K^-}$C 
and by approximately 6~MeV in $^{40}_{K^-}$Ca. The decay width is then 
enhanced by about 6~MeV for carbon and by about 3~MeV for calcium. This enhancement 
of the decay width is a consequence of the $K^-$ binding energy dependence 
of $\Gamma_{K^-}$ in the relevant region of $B_{K^-}$ 
(see Fig.\ \ref{fig:GC}) and also of the moderate increase of the nuclear 
density distributions when compared to the case of purely $s$-wave 
interactions. 
\begin{table}
\caption{
$p$-wave interaction contributions to the $K^-$ binding energy $B_{K^-}$, 
to the single-particle binding energy $B_{K^-}^{\rm sp}$ and to the width 
$\Gamma_{K^-}$, for the $1s$ $K^-$--nuclear states in $^{12}_{K^-}$C and 
in $^{40}_{K^-}$Ca, using the NL-SH parameterization. Results for $s$-wave 
interactions exclusively are denoted by S and results including the $p$-wave 
interaction Eqs.~(\protect{\ref{eq:pwave}}) and (\protect{\ref{eq:c0}}) 
are denoted by S+P.}
\label{tab:1s}
\begin{ruledtabular}
\begin{tabular}{lcccccc}
 &\multicolumn{3}{c}{$^{12}$C} &  \multicolumn{3}{c}{ $^{40}$Ca} \\
 & $B_{K^-}$ (MeV) & $B_{K^-}^{\rm sp}$ (MeV) & $\Gamma_{K^-}$ (MeV) &
$B_{K^-}$ (MeV) & $B_{K^-}^{\rm sp}$ (MeV) & $\Gamma_{K^-}$ (MeV)
\\
\hline
S   & 100.0 & 109.8 & 51.1& 100.0 & 104.4 & 35.0 \\
 S+P  & 112.8 & 123.3 & 56.9 & 105.6 & 111.8 & 38.3 \\
\end{tabular}
\end{ruledtabular}
\end{table}

\subsection{Multi-$\bar K$ nuclei} 
\label{sec:multi}
In the second part of this work, we embedded several ($\kappa \geq 2$) 
antikaons in the nuclear medium and studied the nuclear response, 
as well as the energies and widths of bound states in such multi-$\bar 
K$ nuclear systems. We studied nuclear systems containing only $K^-$ 
mesons or only ${\bar K}^0$ mesons. 

\begin{figure} 
\includegraphics[width=14cm]{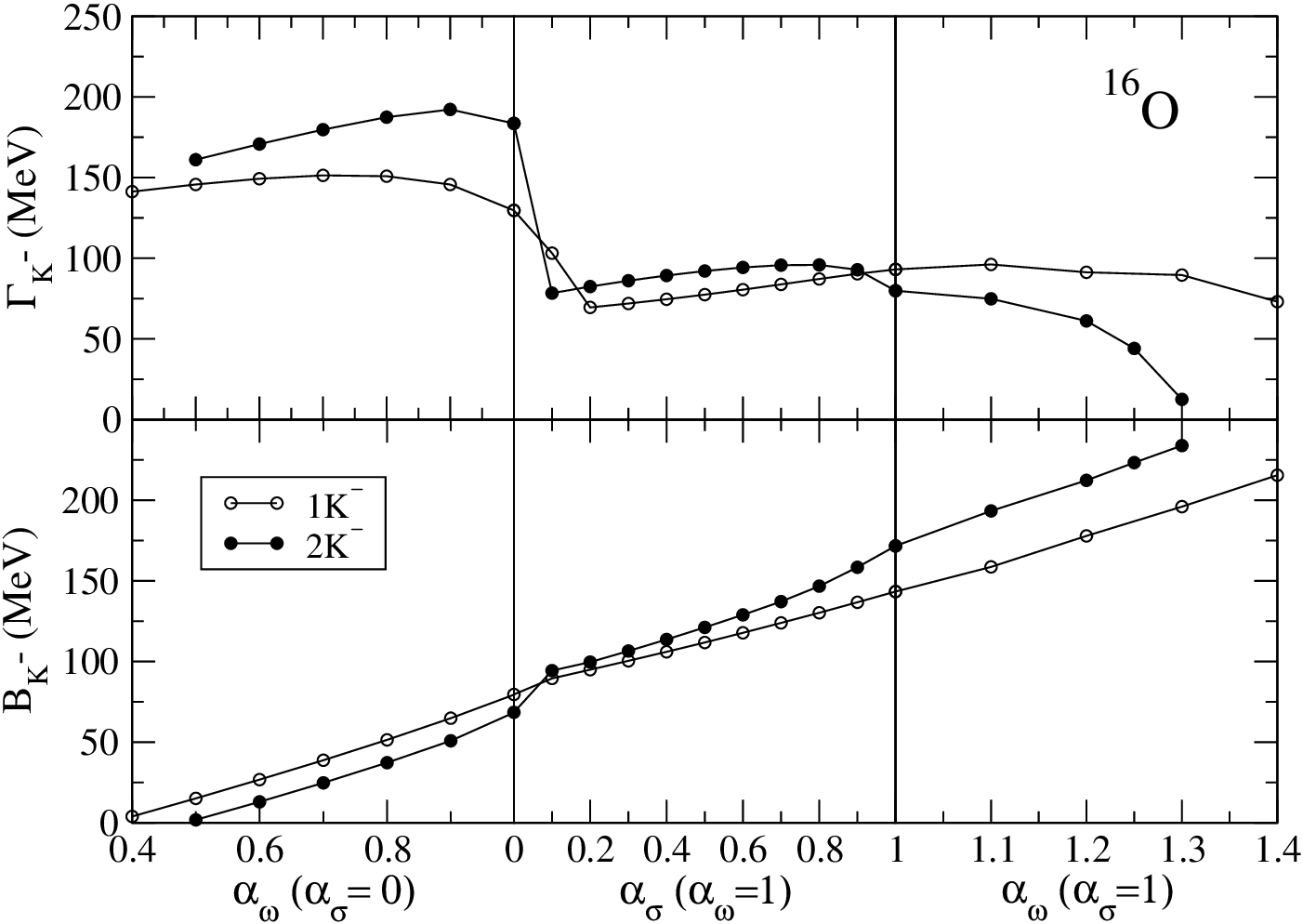} 
\caption{$1s$ $K^-$ binding energy (bottom panels) and width (top panels) in 
$^{16}$O with one and two antikaon(s) as a function of the coupling strengths 
$\alpha_\omega$ and $\alpha_\sigma$ (see text), using the NL-SH RMF model.} 
\label{fig:GBO}
\end{figure}

\begin{figure}
\includegraphics[width=12cm]{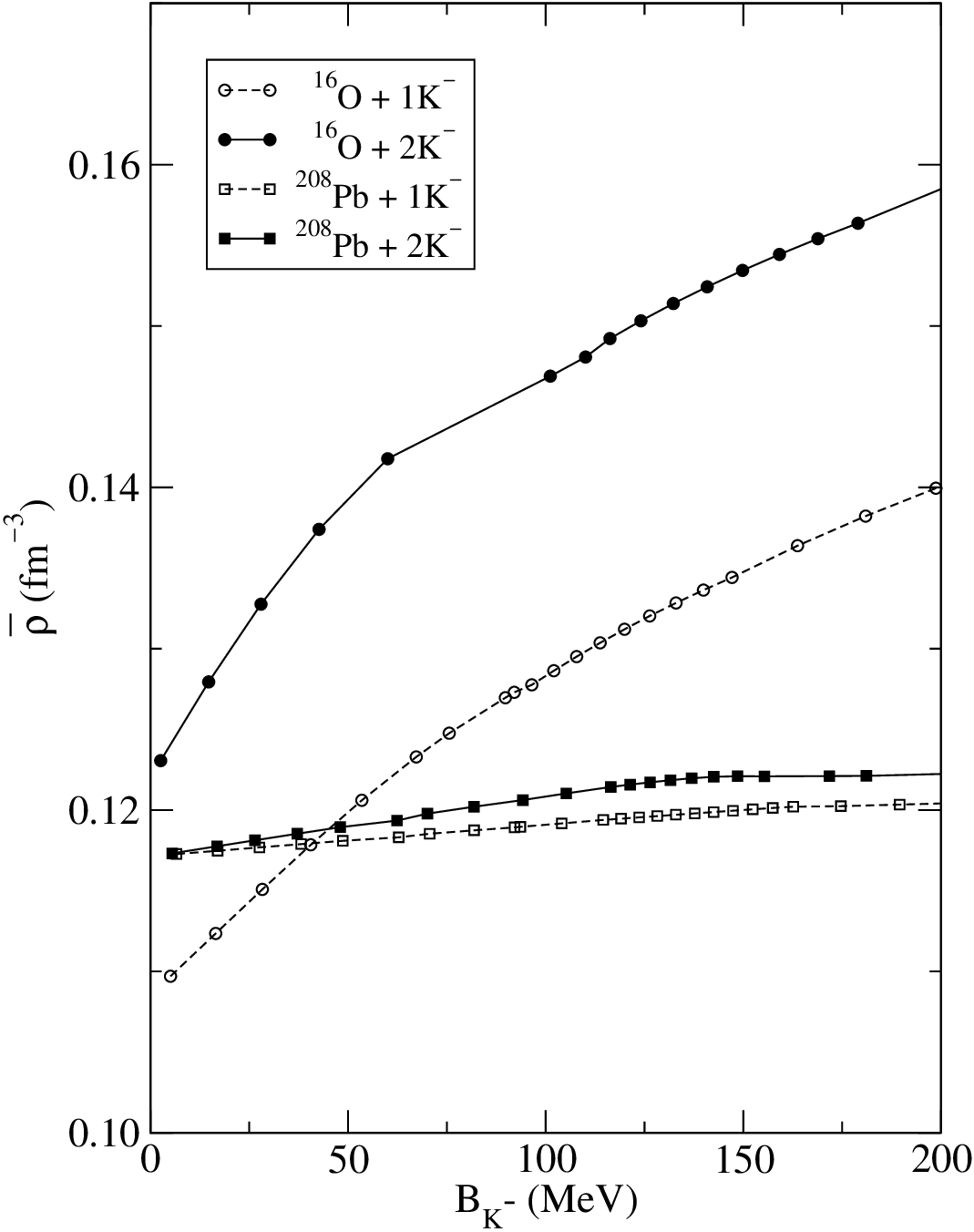} 
\caption{Average nuclear density $\bar \rho$ for $^{16}$O and $^{208}$Pb 
with one and two antikaon(s) as a function of the $1s$ $K^-$ binding energy, 
using the NL-SH RMF model.}
\label{fig:rhob}
\end{figure}

Figure~\ref{fig:GBO} shows the calculated binding energies and widths of 
$1s$ $K^-$ states in $^{16}$O with two bound antikaons, using the NL-SH 
model, in comparison to similar calculations for a single antikaon bound 
in $^{16}$O. The $K^-$ binding energy $B_{K^-}$ of the second $K^-$ in the 
double-$K^-$ nucleus $^{16}_{2K^-}$O is lower than the $K^-$ binding energy 
in $^{16}_{1K^-}$O for binding energies $B_{K^-}$ $\lesssim 90$~MeV. 
Primarily, this is a consequence of the dominance of the mutual repulsion 
induced by the vector-meson mean fields between the two $K^-$ mesons over 
the extra polarization of the nuclear core evoked by the presence of the 
second $K^-$ meson. It is worth noting that this result is amplified by the 
larger width $\Gamma_{K^-}$ in the case of two antikaons, which acts 
repulsively, and by setting $\alpha_{\sigma}=0$, for the attractive 
interaction generated by the scalar mean field, at the low $B_{K^-}$ region 
of the left-hand panels in the figure. (If the coupling of the $K^-$ meson 
to {\it all} the vector-meson mean fields is switched off, so that $B_{K^-}$ 
is generated solely via the scalar $\sigma$-meson mean field, and furthermore 
the imaginary potential is switched off, the binding energy $B_{K^-}$ of the 
second $K^-$ in $^{16}_{2K^-}$O is {\it always} larger than $B_{K^-}$ in 
$^{16}_{1K^-}$O.) This hierarchy is reversed at 
$B_{K^-}\simeq 90$ MeV when the $K^-$ binding energy in $^{16}_{2K^-}$O 
becomes larger than in $^{16}_{1K^-}$O, reflecting a strong polarization of 
the nuclear core (see Figs.\ \ref{fig:rhob} and \ref{fig:rhoO}). 
The enhancement of the binding energy $B_{K^-}$ in the double $K^-$ nucleus 
is then responsible for the crossings of the curves for the $K^-$ decay 
widths $\Gamma_{K^-}$, at $B_{K^-}\simeq 90$ and 170~MeV, caused by the 
binding energy dependence of the suppression factors. Finally, the sharp 
decrease of the width $\Gamma_{K^-}$ in $^{16}_{2K^-}$O, at 
$B_{K^-}\simeq 230$~MeV, is due to the disappearance of the $2N$-absorption 
channel $\bar K NN \to \Sigma N$. 

Figure~\ref{fig:rhob} shows the average nuclear density $\bar \rho$ in 
$^{16}$O and in $^{208}$Pb with one and with two $K^-$ mesons as a function of 
the $K^-$ binding energy. Adding the second $K^-$ to the nuclear system leads 
to further polarization of the nuclear core. The enhancement of the average 
nuclear density is quite pronounced in light nuclei ($^{16}$O) while in 
heavy nuclei ($^{208}$Pb) it is rather weak. 

\begin{figure} 
\includegraphics[width=12cm]{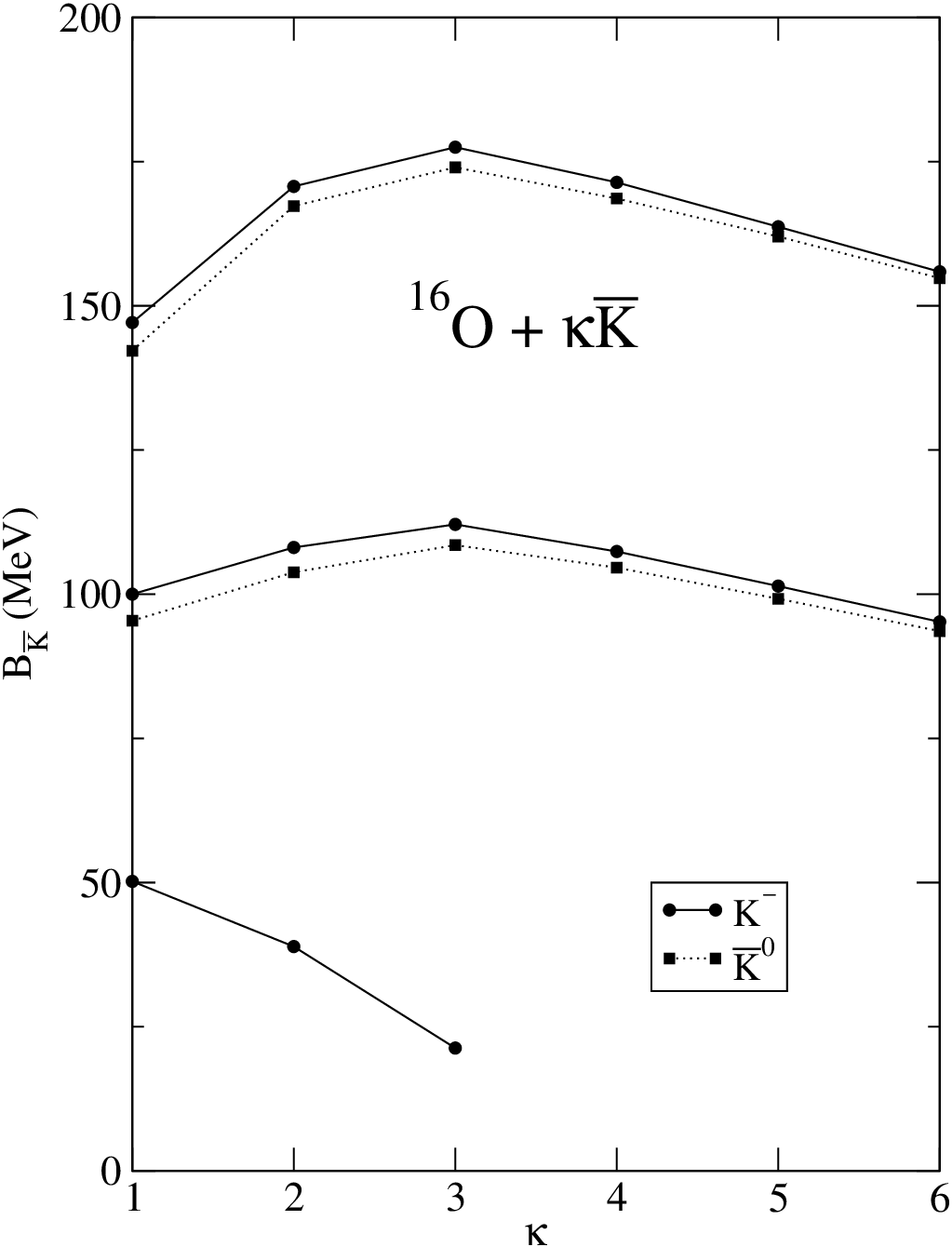} 
\caption{1s $\bar K$ binding energy $B_{\bar K}$ in $^{16}{\rm O}+ \kappa 
{\bar K}$, where $\bar K = K^-$ (circles) or ${\bar K}^0$ (squares), as a 
function of the number $\kappa$ of antikaons, using the NL-SH RMF model.} 
\label{fig:Okappa}
\end{figure}

Figure~\ref{fig:Okappa} presents the 1s $\bar K$ binding energy $B_{\bar K}$ 
in the multi-$\bar K$ nuclei $^{16}{\rm O}+\kappa {\bar K}$, where 
$\kappa {\bar K} = \kappa K^-$ or $\kappa {\bar K^0}$, as a function of the 
number of antikaons $\kappa$, calculated within the NL-SH RMF 
parameterization. The figure demonstrates that increasing the number 
of antikaons in the nuclear medium does not necessarily lead to a sizable 
increase of the binding energy $B_{\bar K}$. Just on the contrary, for 
relatively small values of $B_{\bar K}$ (the curve starting with 50 MeV for 
$\kappa = 1$ in the figure), $B_{\bar K}$ decreases as a function of $\kappa$. 
This is consistent with the trend shown in Fig.~\ref{fig:GBO} for one and 
two $K^-$ mesons. (Had we replaced the vector-meson mean-field couplings 
by an equivalent purely scalar-meson mean-field coupling to yield the same 
starting value of $B_{\bar K}$, setting also Im~$\Pi_{\bar K}=0$, 
$B_{\bar K}$ would rather increase as a function of $\kappa$.) 
For the higher starting values for $B_{\bar K}$ in the figure, a moderate 
decrease of $B_{\bar K}$ as a function of $\kappa$ occurs for $\kappa > 3$, 
indicating that the ${\bar K}$ binding energies have reached saturation. 
We note that the difference between the $K^-$ and ${\bar K}^0$ curves is 
relatively small, a few MeV at most, decreasing with $\kappa$ owing to the 
increased role of the Coulomb repulsion among the $K^-$ mesons. 

\begin{figure} 
\includegraphics[width=12cm]{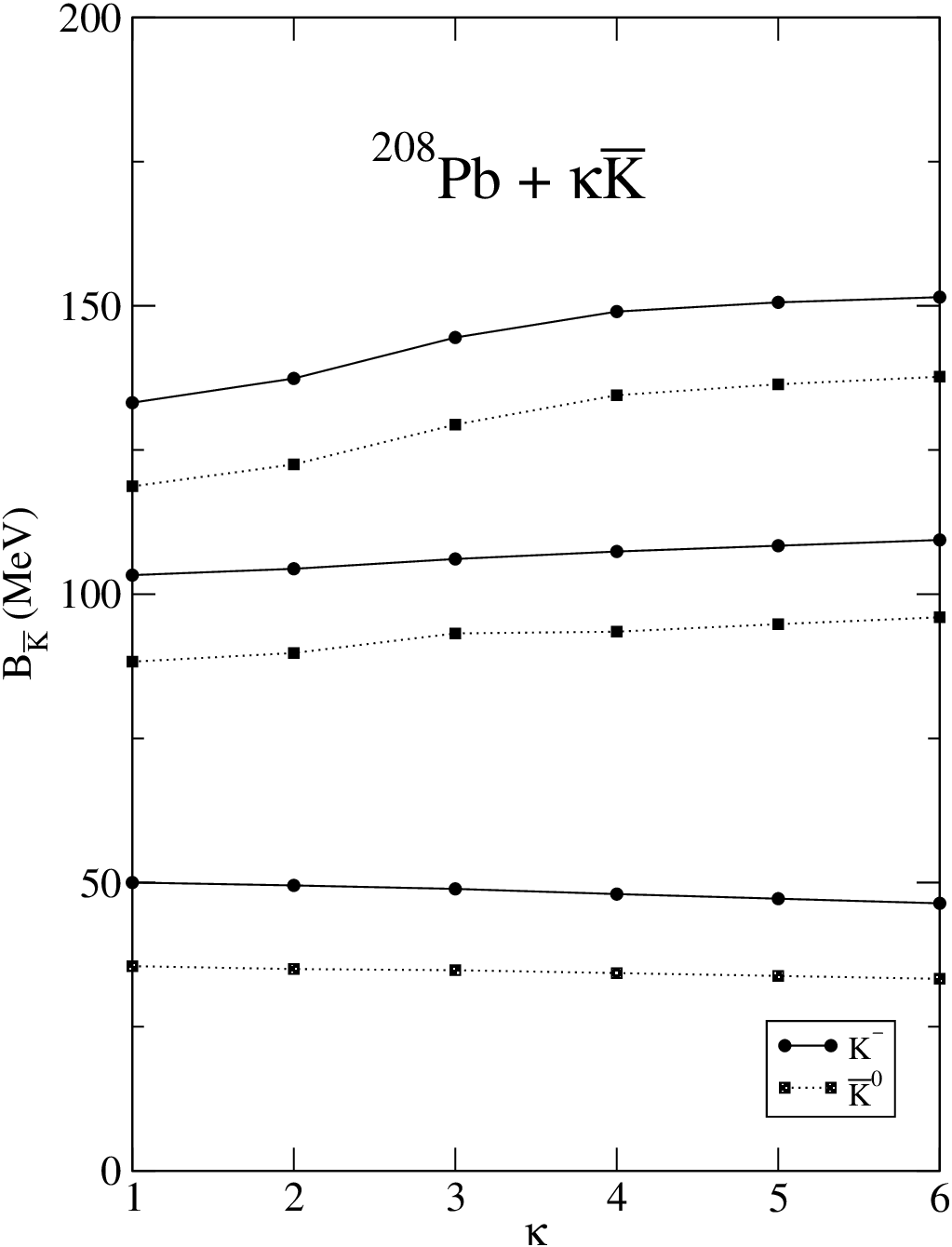} 
\caption{Same as Fig.\ \protect{\ref{fig:Okappa}}, but for 
$^{208}{\rm Pb} + \kappa {\bar K}$, where ${\bar K} = K^-$ (circles) 
and ${\bar K}^0$ (squares).} 
\label{fig:Pbkappa}
\end{figure}

\begin{figure} 
\includegraphics[width=12cm]{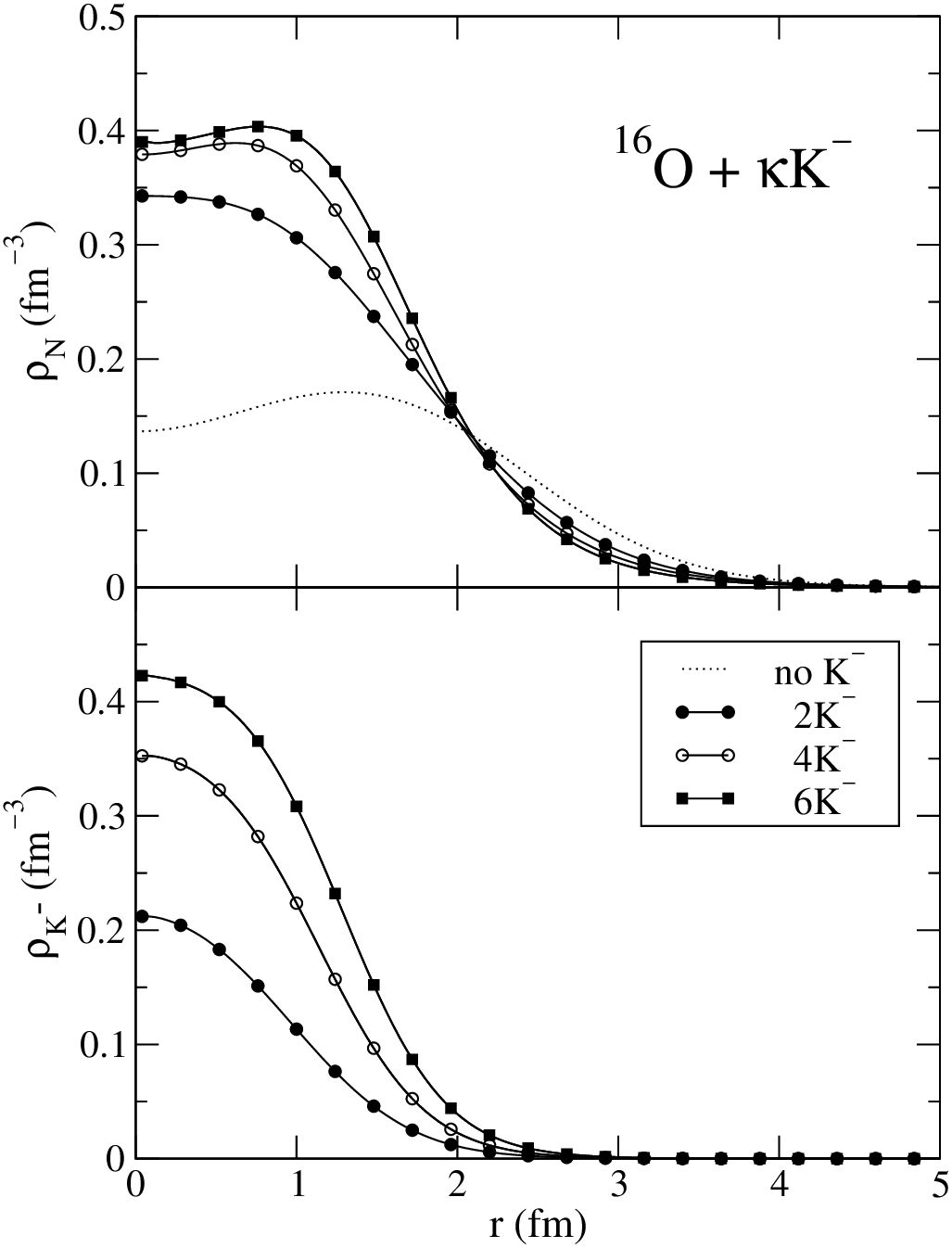} 
\caption{
Nuclear density $\rho_N$ (top panel) and $1s$ $K^-$ density 
$\rho_{K^-}$ (bottom panel) in $^{16}{\rm O}+ \kappa K^-$, 
using the NL-SH RMF model with 
$\alpha_{\sigma}=0.26$ and  $\alpha_{\omega}=1$, yielding $B_{K^-}=100$~MeV 
in $^{16}{\rm O} + 1K^-$. The dotted curve stands for the $^{16}$O 
density in the absence of $K^-$ mesons.} 
\label{fig:rhoO} 
\end{figure}

Figure~\ref{fig:Pbkappa} shows the 1s $\bar K$ binding energy in the 
multi-${\bar K}$ nuclei $^{208}{\rm Pb}+\kappa {\bar K}$, where 
$\kappa {\bar K} = \kappa K^-$ or $\kappa {\bar K^0}$, calculated using 
the NL-SH model. It is found that the attractive Coulomb interaction of 
a $K^-$ meson with the large number of protons ($Z=82$ for $^{208}{\rm Pb}$) 
outweighs the repulsion due to the $\rho{\bar K}$ coupling, so that 
the lowest-energy configuration is provided by a purely $K^-$ charge 
configuration. The $K^-$ curves are displaced by about 15 MeV to higher 
values of binding energies than the respective ${\bar K}^0$ curves. 
Here and in the previous figure, the ${\bar K}^0$ couplings to the 
isoscalar-meson mean fields were taken identical to the corresponding 
$K^-$ couplings, while differing in sign for the isovector $\rho$-meson 
mean field. Furthermore, Im~$\Pi_{\bar K}$ was assumed to be the same 
for ${\bar K}^0$ mesons as for $K^-$ mesons. 
For the lowest starting value of $B_{\bar K}$ in Fig.~\ref{fig:Pbkappa}, 
as for $^{16}$O in the previous figure, $B_{\bar K}$ decreases 
as a function of $\kappa$, although at a slower rate. 
For higher starting values of $B_{\bar K}$, a moderate increase of 
$B_{\bar K}(\kappa)$ is observed, which gradually slows down with 
increasing the number of antikaons $\kappa$. We checked that saturation 
of $B_{\bar K}(\kappa)$ is actually reached for a higher value of 
$\kappa$ than shown in the figure ($\kappa \geq 10$). 

\begin{figure} 
\includegraphics[width=12cm]{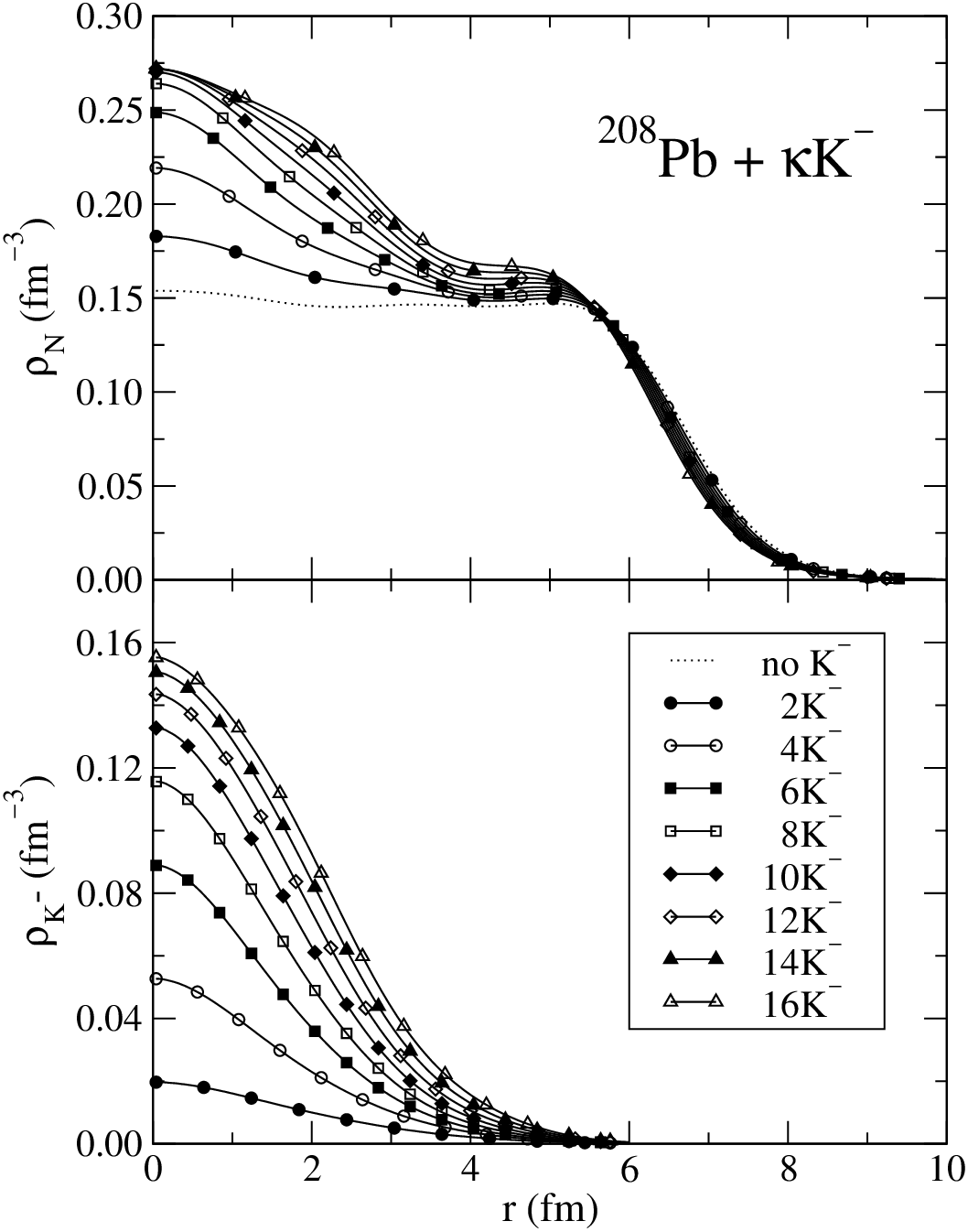} 
\caption{
Same as in Fig.\ \protect{\ref{fig:rhoO}}, but for $^{208}{\rm Pb} + 
\kappa K^-$ with $\alpha_{\sigma}=0$ and $\alpha_{\omega}=0.86$, yielding 
$B_{K^-}=100$~MeV in $^{208}{\rm Pb} + 1K^-$.} 
\label{fig:rhoPb} 
\end{figure}

The dependence of the nuclear density $\rho_N(r)$ and the $K^-$ 
density $\rho_{K^-}(r)$ on the number of $K^-$ mesons 
embedded in the nuclear medium is shown in Fig.\ \ref{fig:rhoO} for 
($^{16}{\rm O} + \kappa  K^-$) and in Fig.\ \ref{fig:rhoPb} for 
($^{208}{\rm Pb} + \kappa K^-$). Shown also for comparison, in dotted lines, 
are the density distributions $\rho_N$ for $\kappa=0$. The $K^-$ couplings 
were chosen such that the 1$K^-$ configuration was bound by 100~MeV. 
The density distributions behave quite regularly as a function of $\kappa$. 
In $^{16}{\rm O} + \kappa K^-$, for $\kappa \geq 4$, the nuclear 
density distribution recovers the saddle it had without antikaons at 
$r \approx 0$. The increase of the nuclear density resulting from the $1s$ 
antikaons is limited to the vicinity of the nuclear center where the 
density $\rho_{K^-}$ is substantial, much the same as for single-$K^-$ 
nuclei (see Fig.~\ref{fig:CaC}) upon increasing $B_{K^-}$. The gradual 
increase of $\rho_{K^-}$ as well as of $\rho_N$ slows down with $\kappa$, 
leading to saturation as demonstrated for $\rho_N(0)$ in 
Figs.\ \ref{fig:rhoO} and \ref{fig:rhoPb}. The saturation is also apparent 
when the $K^-$ densities $\rho_{K^-}$ (normalized to $\kappa$) are shown as 
$\rho_{K^-}(0)/\kappa$ in Table~\ref{tab:2}. 

\begin{table} 
\caption{Values of $\rho_{K^-}(0)/\kappa$ (in fm$^{-3}$) as a function of the 
number of $K^-$ mesons $\kappa$ in $^{16}{\rm O}$ and $^{208}{\rm Pb}$, using 
the NL-SH RMF parameterization. The $K^-$ density $\rho_{K^-}$ is normalized 
to $\kappa$. The $K^-$ coupling constants for each core nucleus give rise to $B_{K^-}=100$ 
MeV for a single $K^-$ meson.}
\label{tab:2}
\begin{ruledtabular}
\begin{tabular}{lccccccc}
 & $\kappa$ & 1 & 2 & 4 & 6 & 8 & 10 \\
\hline
$\rho_{K^-}(0)/\kappa$ & $^{16}{\rm O} + \kappa K^-$ & 
0.098 & 0.106 & 0.088 & 0.070 &  &  \\
$\rho_{K^-}(0)/\kappa$ & $^{208}{\rm Pb} + \kappa K^-$ & 
0.009 & 0.010 & 0.013 & 0.014 & 0.015 & 0.013
\\
\end{tabular}
\end{ruledtabular}
\end{table}

\section{Conclusions} 
\label{sec:concl} 

In the present work, we studied in detail the interplay between the 
underlying dynamical processes and the relevant kinematical conditions which 
determine the decay width of deeply bound $\bar K$--nuclear states in the nuclear 
medium. We performed fully self-consistent dynamical calculations of 
$\bar K$--nuclear states for nuclear systems with one and several $\bar K$ mesons 
within the RMF approach. 

We verified that the interaction of a $1s$ $K^-$ meson with the $\rho$-meson 
mean field affects negligibly the $K^-$ binding energy. Its main effect on the 
nucleon single particle energies is to partly cancel, and for the $1s$ nucleon level even 
reverse the $p-n$ Coulomb energy difference. For all nuclei and RMF 
parameterizations considered in the present work, the $\rho$-meson 
contribution slightly decreases the $K^-$ binding energy $B_{K^-}$ by less than about 
5~MeV for $B_{K^-} \lesssim 200$~MeV. Similarly, the $\phi$-meson 
contribution in systems with several $K^-$ mesons reduces the $K^-$ binding 
energy by a few MeV in this range of $B_{K^-}$ values. 

The calculations involving the $p$-wave interaction of the $K^-$ meson with 
a nucleus indicate that the $p$-wave interaction plays a secondary role for 
deeply bound $K^-$--nuclear systems where the mean-field concept is 
acceptable. Although the $p$-wave interaction by itself is too weak to 
cause nuclear binding, its contribution in the lightest (carbon) nucleus 
considered in the present work amounts to more than 10 MeV and is certainly 
nonnegligible. Since the effect of the $p$-wave interaction appears to 
increase upon decreasing the atomic number, it could play a primary role 
in deeply and tightly bound few-body $K^-$ systems. 

We found that the implementation of the $\pi\Lambda$ decay channel in the 
single-nucleon absorption mode enhances the $K^-$ conversion width for $K^-$ 
binding energies $B_{K^-}\lesssim 170$~MeV. This enhancement is almost uniform 
for both linear and nonlinear parameterizations in all nuclei 
considered. The most remarkable contribution occurs for $K^-$ binding 
energies in the range $B_{K^-} \approx 100-160$~MeV where it reaches values of 
approximately 20~MeV. The assumption of a $\rho^2$ density dependence for the 
$2N$-absorption mode adds further conversion width especially to the deeply 
bound $K^-$--nuclear states. The increase is particularly large for nonlinear 
parameterizations owing to the strong polarization effects affordable through 
the moderate value of nuclear incompressibility, as opposed to the highly 
unrealistic values in linear parameterizations. Altogether, the results of 
these comprehensive calculations suggest that $K^-$ total decay widths 
for deeply bound $K^-$ nuclear states ($B_{K^-} > 100$~MeV) are substantial, 
$\Gamma_{K^-} \sim 50 - 100$~MeV. 

We also studied nuclear systems containing several antikaons. The nuclear and 
$\bar K$ densities were found to behave quite regularly upon increasing the 
number of antikaons embedded in the nuclear medium. The calculations do not 
indicate any abrupt or substantial increase of the densities. 
The central nuclear densities in multi-$K^-$ $^{16}{\rm O}$ nuclei and in 
multi-$K^-$ $^{208}{\rm Pb}$ nuclei appear to saturate at about only 50\% and 
60\%, respectively, higher values than the central nuclear densities in 
the corresponding systems with one antikaon. Furthermore, the $\bar K$ 
binding energy saturates upon increasing the number of $\bar K$ mesons 
embedded in the nuclear medium. The heavier the nucleus is, the more 
antikaons it takes to saturate the binding energies, but even for 
$^{208}{\rm Pb}$ the number required does not exceed approximately 10. 
The saturated values of $\bar K$ binding energies do not exceed the range 
of values 100--200~MeV considered normally as providing deep binding for 
one antikaon. This range of binding energies leaves antikaons in 
multi-${\bar K}$ nuclei comfortably above the range of energies where 
hyperons might be relevant. It is therefore unlikely that multi-${\bar K}$ 
nuclei may offer precursor phenomena in nuclear matter towards kaon 
condensation. This does not rule out that kaon condensation occurs in 
neutron stars where different constraints hold for the composition of matter. 
Although we presented results for one particular choice of RMF model, 
the NL-SH model \cite{SNR93}, the use of other realistic mean-field models 
supports these conclusions. 
\begin{acknowledgments}
One of us (AG) acknowledges and thanks Wolfram Weise for stimulating 
discussions, particularly on the $\bar{K}$--nuclear $p$-wave interaction. 
This work was supported in part by the GA AVCR grant IAA100480617
and by the Israel Science Foundation grant 757/05.
\end{acknowledgments}

\newpage

\end{document}